\documentclass[letterpaper, 10 pt, conference]{ieeeconf}  

\IEEEoverridecommandlockouts                              

\usepackage{graphics} 
\usepackage{epsfig} 
\usepackage{mathptmx} 
\usepackage{times} 
\usepackage{amsmath} 
\usepackage{amssymb}  
\usepackage[ruled,vlined]{algorithm2e}
\usepackage{multirow}
\usepackage{color}
\usepackage{url}
\usepackage{booktabs}

\usepackage{hyperref}
\usepackage{graphicx}       
\usepackage{subcaption}
\usepackage[bottom]{footmisc}

\title{\LARGE \bf
Centralizing State-Values in Dueling Networks for \\Multi-Robot Reinforcement Learning Mapless Navigation
}

\author{Enrico Marchesini$^{*}$, Alessandro Farinelli
\thanks{$^{*}$Contact author: {\tt\small enrico.marchesini@univr.it}}%
\thanks{Authors are with the Department of Computer Science, University of Verona, 37135 Verona, Italy.}%
}

\begin{document}

\maketitle
\thispagestyle{empty}
\pagestyle{empty}


\begin{abstract}
We study the problem of multi-robot mapless navigation in the popular Centralized Training and Decentralized Execution (CTDE) paradigm. This problem is challenging when each robot considers its path without explicitly sharing observations with other robots and can lead to non-stationary issues in Deep Reinforcement Learning (DRL). The typical CTDE algorithm factorizes the joint action-value function into individual ones, to favor cooperation and achieve decentralized execution. Such factorization involves constraints (e.g., monotonicity) that limit the emergence of novel behaviors in an individual as each agent is trained starting from a joint action-value. In contrast, we propose a novel architecture for CTDE that uses a centralized state-value network to compute a joint state-value, which is used to inject global state information in the value-based updates of the agents. Consequently, each model computes its gradient update for the weights, considering the overall state of the environment. Our idea follows the insights of Dueling Networks as a separate estimation of the joint state-value has both the advantage of improving sample efficiency, while providing each robot information whether the global state is (or is not) valuable. Experiments in a robotic navigation task with 2 4, and 8 robots, confirm the superior performance of our approach over prior CTDE methods (e.g., VDN, QMIX).
\end{abstract}
\section{Introduction}
\label{sec:introduction}

Multi-Agent Reinforcement Learning (MARL) \cite{marl_coopeartive} considers multiple learning agents (or robots) that have to optimize either an individual or a joint reward signal accumulated over time. In particular, we are interested in Multi-Robot navigation as it recently gained attention due to the wide range of applicability (e.g., multi-robot search \cite{mr_search}).

Multi-robot navigation has been previously considered as a cooperative MARL problem and requires the learning of an efficient collision avoidance policy \cite{mr_collisionavoidance}. Among the different approaches that can address MARL, \textit{centralized learning} aims at reducing the task to a single-robot problem that can be solved with standard DRL algorithms. However, these centralized solutions require comprehensive knowledge about all the robots' observations (e.g., positions, velocities) and their workspace (e.g., a grid map), which are concatenated to form a joint observation. Intuitively, centralized methods do not scale well within the number of robot and consistently fail in practice due to the "lazy" agent problem \cite{vdn} (i.e., an individual that is discouraged from learning the policy as its exploration can hinder other successful agents). 

Conversely, \textit{independent learning} consists of individual agents that have to face a non-stationary problem, due to the dynamics of the environment that changes within the other unobserved policies. In practice, the non-stationarity can be addressed by using recurrent units (e.g., Long Short-Term Memory), but the lack of shared information does not favor cooperative behaviors. However, independent learning is typically used as a baseline as it represents a simple yet good-performing solution for a variety of MARL tasks \cite{maddpg}.

    \begin{figure}[b]	
    \begin{center}
    	\includegraphics[width=.9\linewidth]{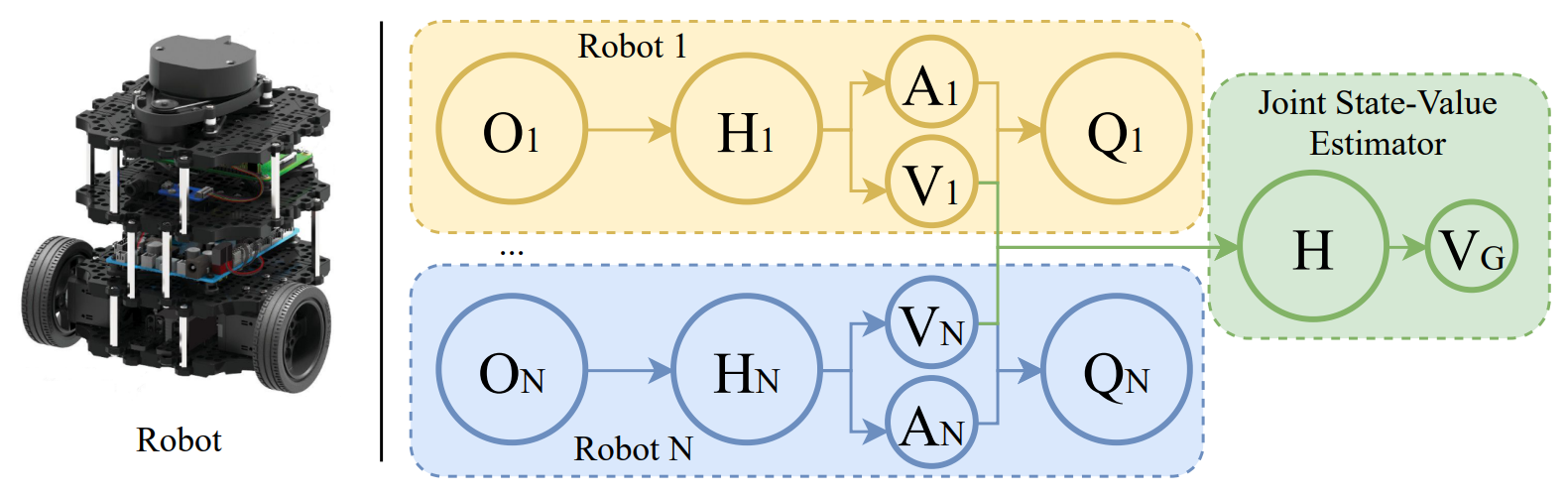}
        \caption{Left: Robotic platform. Right: Architecture of GDQ. Each robot uses local observation $O_i$ (and hidden layers $H_i$) to compute the $Q_i$-values for a decentralized execution. Individuals Dueling Architecture compute the state-values $V_i$ that are used as input for the joint state-value estimator.}
        \label{fig:overall}
    \end{center}
    \end{figure}
    
Both the scalability and the non-stationarity issues have been considered by the recent \textit{Centralized Training with Decentralized Execution} paradigm, which has been used to design value-based MARL algorithms \cite{vdn, qmix, qtran} that showed state-of-the-art results on several multi-agent tasks (e.g., multi-agent particle envs \cite{particleenv}). CTDE involves the training of robots' policies that use global information in a centralized way and rely only on local observations for the action selection, hence enabling the decentralized execution. In more detail, to address the combinatorial action space and the dimensionality of the joint observations, VDN \cite{vdn}, QMIX \cite{qmix}, and QTRAN \cite{qtran} propose different mechanisms to factorize the joint (or global, interchangeably) action-value function, which we refer to as $Q_{G}$ for notation. However these approaches suffer from a common limitation as they introduce structural constraints (e.g., additivity and monotonicity) to ensure the factorization of $Q_{G}$ into the agent individuals $Q_{i}$-values (with $i \in [1, .., N]$, where $N$ is the number of robots) that are used for the decentralized action selection process. Such constraints severely limit the joint action-value function class that VDN and QMIX can represent, while QTRAN showed poor practical performance despite its strong theoretical factorization guarantees \cite{qtran}.

To address these limitations, we present a novel value-based CTDE approach (Fig. \ref{fig:overall}), called \textit{multi-agent Global Dueling Q-learning} (GDQ), that in contrast to prior work exploits the state-values $V$\footnote{We omit the state $s$ and action $a$ parameters for notation simplicity.} of each agent to estimate a joint state-value for the system, which is used in the update of the agents (as detailed in Section \ref{sec:methods}). In more detail, GDQ centralizes the individuals state-values of the agents, that are estimated using Dueling Networks \cite{duelingdqn}, to compute a joint state-value $V_{G}$ with a Double Deep Q-Network (DDQN) \cite{doubledqn} (which is trained using the cumulative reward signal). Our goal is to avoid the issues of factorizing $Q_G$ by exploiting the separate state-value $V$ and advantage $A$ functions of each agent. As showed by Dueling Networks, the combination of $V$ and $A$ results in the agent's action-value $Q$, from which the action of an agent is sampled. Intuitively, the state-value function measures how good it is to be in a state, whether the advantages are the relative importance of the actions. The combination of these values $Q = V + A$ is then used to measure the value of choosing a particular action in the state. Hence, we use the agents' state-values as input for the centralized DDQN, which is used to learn a joint state-value $V_{G}$. This $V_{G}$ replaces $V$ in the estimation of the temporal difference target of each agent, and also has the advantage of addressing the overestimation problem of Deep Q-Network-based algorithms that typically requires a target network \cite{doubledqn}. Our idea is to inject information on the global environment state in each agent update rule, such that the robots optimize their behaviors while considering the state of the system.

We first evaluate the performance of GDQ on the explanatory \textit{Cooperative navigation} task of the multi-agent particle envs benchmark suite \cite{particleenv}, to confirm the superior performance of GDQ with respect to prior value-based CTDE algorithms (i.e., VDN and QMIX). Hence, we present a Turtlebot3\footnote{https://www.turtlebot.com/} multi-agent navigation task with 2, 4, and 8 robots, to highlight the scalability and the beneficial effects of GDQ in increasingly complex collision avoidance situations.

\section{Background and Related Work}
\label{sec:background_related}

Similar to previous CTDE approaches, we can describe our multi-robot navigation scenario as a Dec-POMDP \cite{decpomdp} (we remind the interested reader to prior work for details about the Dec-POMDP formalization \cite{qmix}, \cite{eumaspomcp}). Given the partial observability setting of the tasks, the $Q$-functions are defined over individual observation histories, hence we incorporate recurrent units (i.e., an LSTM layer) in the robots' networks. For each agent, we also consider a common setup with Prioritized buffers \cite{per} and n-step returns \cite{rainbow} to speed up the training.

\subsection{Deep Reinforcement Learning for Robotic Navigation}
Robotic navigation is a well-known problem in recent DRL literature \cite{drl_navigation1, drl_navigation3, prove, superl}. This class of problems has been naturally extended to the MARL domain, due to the wide range of multi-robot navigation applications, such as search and rescue \cite{mr_search}, or collision avoidance solutions \cite{mr_collisionavoidance}. Typical solutions to these problems consider policy-gradient algorithms with continuous high-dimensional action spaces. However, recent work \cite{drl_navigation2, mixed_gdrl} show that value-based methods with discrete action spaces achieve similar performance while drastically reducing training time. Furthermore, several different methods are renewing the interest in the use of discrete action spaces, showing that value-based DRL algorithms can also handle high-dimensional actions. In more detail, the Branching Architecture \cite{branching} presents an adaptation of the Dueling Network (that we consider in our work) and obtain competitive results over continuous algorithms in classical MuJoCo locomotion tasks \cite{mujoco}, with several other value-based methods that confirm these results \cite{valuebased1, valuebased2}. Hence, similarly to the CTDE baselines \cite{vdn, qmix}, we use discrete action spaces for our tasks in Section \ref{sec:methods}, \ref{sec:evaluation}.

\subsection{Centralized Training with Decentralized Execution}
CTDE has recently attracted attention as multi-agent DRL \cite{vdn, qmix, qtran} paradigm. The main idea introduced in one of the first DRL CTDE approach, the actor-critic MADDPG \cite{maddpg}, is that the individuals are trained using centralized information (e.g., $Q_G$) or the global state (e.g., positions and goals of the other robots) during the training phase. However, the decision-making process of each agent is strictly independent as it is based only on the local action-observation history, guaranteeing a decentralized execution. 

Since MADDPG (and the actor-critic COMA \cite{coma}), the main research trend is to develop CTDE value-based approaches, that aim at satisfying the \textit{Individual-Global-Max} (IGM) requirement, which state that the optimal joint action induced from $Q_G$ is equivalent to the collection of the individuals $Q_i$-values. Hence, this asserts that $Q_G(\mathbf{\tau}, \mathbf{a})$ (where $\mathbf{\tau}$ is the joint action-observation history, and $\mathbf{a} \equiv [a_i]^{N}_{i=1}$ is the joint action) is factorizable if and only if exists $[Q_i : \mathbf{T} \times \mathbf{A} \rightarrow \mathbb{R}]^{N}_{i=1}$ such that $\forall \mathbf{\tau} \in \mathbf{T}$:

\begin{equation}
    \operatorname*{arg\,max}_{\mathbf{a} \in \mathbf{A}} Q_G(\mathbf{\tau}, \mathbf{a}) = 
    \left( \begin{array}{ccc}
   \operatorname*{arg\,max}_{a_1 \in A} Q_1(\tau_1, a_1) \\
    \vdots \\
    \operatorname*{arg\,max}_{a_N \in A} Q_N(\tau_N, a_N)
    \end{array} \right)
    \label{eq:igm}
\end{equation} \newline
\noindent we can summary Eq. \ref{eq:igm} saying that a global $\operatorname*{arg\,max}$ on $Q_G$ returns the same result as a set of $\operatorname*{arg\,max}$ on each $Q_i$ (with $i \in [1, .., N]$).
Two different factorization strategies for satisfying the IGM have been proposed: (i) \textit{additivity} by VDN \cite{vdn}, and (ii) \textit{monotonicity} by QMIX \cite{qmix}. These factorizations are sufficient for the IGM principle. However, they limit the representation expressiveness of the joint action-value function $Q_G$ \cite{poor_factorization}. QMIX also considered additional \textit{hypernetworks} \cite{hypernetworks} to add additional global state information to the robot individual observations, introducing overhead. QTRAN \cite{qtran} uses a relaxation of the IGM constraint in the form of a linear constraint and soft regularization, which despite the theoretical guarantee result in poor performance.

In contrast to this research trend that aims at factorizing $Q_G$, and use additional global state information to favor cooperation, GDQ is more similar to the \textit{independent learning} paradigm, in the sense that the robots do not share any additional global observation among each other (to favor cooperation), but they only rely on an additional estimation of a joint state-value $V_G$, which uses the individuals $V$ as input (i.e., it is a form of centralized training). Crucially, $V_G$ is only used in the temporal difference target computation of each agent, hence GDQ is also based on the CTDE paradigm.

\section{Methods}
\label{sec:methods}

In this section, we first introduce our multi-agent Global Dueling Q-learning algorithm in detail. We also discuss the benefits of computing a joint state-value with respect to prior CTDE approaches that factorize the joint action-value to satisfy the IGM, which is the main limitation of such methods. We use a multi-agent benchmark, the \textit{Cooperative Navigation} task of the particle envs suite \cite{particleenv} as a motivating example to show the superior performance of our approach.

\subsection{Multi-Agent Global Dueling Q-learning}

GDQ uses a DDQN algorithm for each agent\footnote{Here we refer only to local state/action information, where $s$ could be interpreted as a local observation history in the partial-observable scenario.}, enhanced with priority buffer \cite{per}, n-step returns \cite{rainbow}, and more importantly the dueling architecture \cite{duelingdqn}. The insights of Dueling Networks, in fact, represents the foundation of our value-based CTDE approach. In detail, the agents' $Q$-networks maintain two streams to represent both the individual state-value function $V(s)$, and the individual advantage function $A(s, a)$. The streams are combined with an aggregation layer to produce the agent's action-value function $Q(s, a)$, which in its simplest form computes the $Q$-values as follows:

\begin{equation}
    Q(s, a) = V(s) + A(s, a)
    \label{eq:duelingq}
\end{equation}

The use of a separate representation for the state-value and the action advantages allows to learn whether the considered state $s$ is either valuable or not, and crucially $V(s)$ does not influence the action selection process of an agent, which is strictly dependent on the advantage values:

\begin{equation}
    \operatorname*{arg\,max}_{a} Q(s, a) \equiv \operatorname*{arg\,max}_{a} A(s, a)
\end{equation}

The action-value decomposition improves the sample efficiency of the training process \cite{duelingdqn}. For example, in some states, it is crucial to know which action to take (e.g., to avoid collisions), but in many other states (e.g., where obstacles are far from the agents) the difference between various actions can be negligible. This is mainly related to two factors: 
(i) the direct estimation of the $Q$-values (as in standard Deep Q-Networks) requires to calculate the value of each action at each state, which is a non-negligible overhead when the whole set of possible actions do not affect the environment in a relevant way. For example, in a navigation scenario, we could move to the left or to the right only when there is a risk of collision, otherwise, the choice of action has no major effect on what happens. As detailed by prior work \cite{duelingdqn}, this also aggregates similar actions, which further improves the sample efficiency.
(ii) In a Dueling architecture, where state-value is estimated independently by a stream of the network, $V(s)$ is learned more efficiently as its separate estimation stream is updated with every update of the $Q$-values. 

\noindent Note that Eq. \ref{eq:duelingq} is unidentifiable, in the sense that we can not uniquely recover the state-value and the advantages function from given $Q$-values. For example, if we multiply the state-value and divide the action advantages by the same constant value, we will obtain the same $Q$-values. This causes poor practical performance and has been addressed using different aggregation functions to combine the two streams. Hence, we consider the the best operator identified in the original work \cite{duelingdqn} for the aggregation layer of each agent: 

\begin{equation}
    Q(s, a) = V(s) + \left( A(s, a) - \frac{1}{|A(s, a)|} \sum_{a' \in A(s, a)} a'\right)
    \label{eq:duelingq2}
\end{equation} \newline
\noindent where $|A(s, a)|$ is the cardinality of the advantage function's stream, i.e., the number of possible actions. Given the weights of each agent $\theta_{a_i}$ and the estimated joint next state-value $V_G'$ (described in the next section), GDQ optimizes at iteration $t$ the following loss function for each agent:

\begin{equation}
    L_t(\theta_{a_i}) = \mathbb{E}_{s, a, r, s', V_G'} \left[ \left(y_t(r, s', V_G' \vert \theta_{a_i}) - Q(s, a \vert \theta_{a_i} \right)^2 \right]
    \label{eq:loss}
\end{equation}\newline
\noindent where $s, a, r, s'$ is a tuple of state, action, reward, next state (at iteration $t$) sampled from the agent's memory. In detail, the target is computed by replacing the next state-value, with the one computed by our centralized joint state-value estimator:

\begin{equation}
    \begin{split}
    & y_t(r, s', V_G' \vert \theta_{a_i}) = r + \gamma~ \operatorname*{arg\,max}_{a'}Q(s', a' \vert \theta_{a_i}, V_G') \\
    & Q(s', a' \vert \theta_{a_i}, V_G') = V_G' + A(s', a' \vert \theta_{a_i})
    \end{split}
\end{equation} \newline
As previously discussed, this does not influence the action selection process of an agent as it depends on the advantage function, but it serves to inject learned information on the global state value. The other natural contribution of the introduction of $V_G'$ is that it naturally mitigates the overestimation bias of the standard Deep Q-Networks, and it is possible to remove the requirements of a target network for each agent (as it is typically used to compute $Q(s', a')$ to reduce the overestimation and improve the stability of the training).

\subsection{Joint State-Value Estimator}
We use a standard DDQN algorithm to compute the joint state-value $V_G'$ for the agents. This network takes as input all the agents' next state-values $[V_1(s_t')$, ..., $V_N(s_t')]$ computed by their separate stream to learn an estimation of the joint state-value $V_G'$. The training procedure for this centralized component resembles the standard DDQN algorithm \cite{doubledqn} (which we refer the interested reader for further details), but considers state-values instead of action-values and the cumulative reward of the agents. In more detail, for each environment iteration $t$ we store in a memory buffer the concatenated agents' state-values $\mathbf{v} = [V_1(s_t)$, ..., $V_N(s_t)]$, the next state-values $\mathbf{v'} = [V_1'(s_t')$, ..., $V_N'(s_t')]$ (that are computed with an additional forward step of the agents' networks) and the agents' cumulative rewards $\mathbf{r}$. Hence, given the weights of both the centralized joint state-value estimator $\theta_{c}$ and its target network $\theta_{c'}$ and a sample from the memory, it aims at optimizing the following loss function:

\begin{equation}
    L_t(\theta_{c}) = \mathbb{E}_{\mathbf{v}, \mathbf{v'}, \mathbf{r}} \left[ \left(y_t(\mathbf{r}, \mathbf{v'} \vert \theta_{c'}) - V(\mathbf{v} \vert \theta_{c}) \right)^2 \right]
    \label{eq:loss2}
\end{equation}\newline
\noindent with $V(\mathbf{v} \vert \theta_{c})$ computed by a forward pass of the joint state-value estimator with input $\mathbf{v}$, and target $y_t$ is computed as:

\begin{equation}
    y_t(\mathbf{r}, \mathbf{v'} \vert \theta_{c'}) = \mathbf{r} + \gamma~ V(\mathbf{v'} \vert \theta_{c'}) 
\end{equation} \newline
Besides the benefits of GDQ, the decoupled nature of the state-value stream could also address competitive MARL, using separate joint state-value estimator for each agent group. We intend to explore this direction as future work. 

\subsection{Motivating Example}
\label{subsec:example}
\begin{figure}[b]	
    \begin{center}
    	\includegraphics[width=1\linewidth]{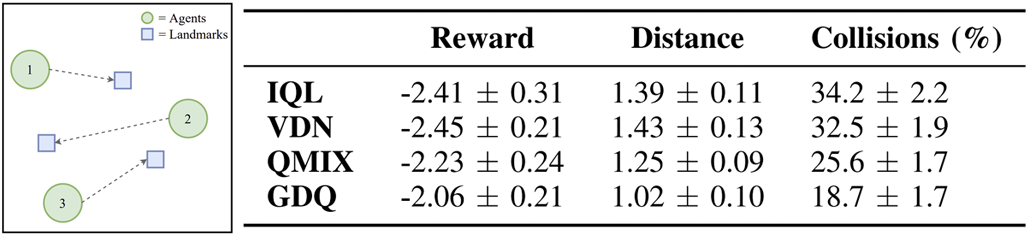}
        \caption{Left: \textit{Cooperative navigation} environment. Right: Average reward, Distance from the landmarks, and Collision percentage for IQL, VDN, QMIX, and GDQ.}
        \label{fig:coopnav}
    \end{center}
    \end{figure}

To evaluate the performance of GDQ over previous similar CTDE baselines (VDN \cite{vdn} and QMIX \cite{qmix}, as QTRAN typically result in poor performance \cite{qtran}), and \textit{independent learners} (IQL) based on DDQN with priority buffer and n-step updates (i.e., similarly to a GDQ agent), we consider the \textit{Cooperative navigation} task of the \textit{multiagent particle envs} suite \cite{particleenv}, as a benchmarking example. All the networks share a similar architecture with 2 hidden layers with 64 \textit{tanh} neurons and Adam optimizer with default learning rate (for algorithmic specific hyper-parameters, we refer the interested reader to the original work of VDN \cite{vdn} and QMIX \cite{qmix} as we used the authors' implementation). In this cooperative problem depicted in Fig. \ref{fig:coopnav} on the left, 3 agents have to cover the same number of landmarks and are rewarded based on how far each one is from each landmark, while they are penalized if they collide with other agents.

Results in Fig. \ref{fig:coopnav} right shows the average reward, average distance from the landmarks, and collision percentage collected in this evaluation over 5 independent runs. Training time where similar for all the considered approaches and was $\approx$100 minutes. These values confirm the superior performance of GDQ over all the considered baselines as it is able to obtain the highest average reward (i.e., minimum average distance), while also favoring information exchange, resulting in the lowest percentage of collisions. Moreover, as detailed in Section \ref{sec:introduction}, IQL obtained good performance despite the simplicity of the approach, while the superior performance of QMIX over VDN and IQL confirm the original authors' results \cite{qmix}.

\section{Evaluation}
\label{sec:evaluation}

Here we introduce our multi-robot mapless navigation for our training and testing experiments with 2, 4, and 8 robots. We then present the results of our evaluation, discussing the benefits of GDQ in the considered robotic navigation tasks.

\subsection{Mapless Navigation Environment Description}

Mapless navigation is a well-known benchmark in recent DRL literature \cite{drl_navigation1, drl_navigation2, drl_navigation3}, and aims at navigating a robot towards a random target using only local observation and the target position, without a map of the surrounding environment or obstacles. Given the wide range of real-world task that involves robot navigation, this problem has been naturally extended to the multi-robot domain \cite{mr_search, mr_selfdriving}. In this paper, we consider a setup similar to the previous single-robot navigation scenarios \cite{drl_navigation2, drl_navigation3}, adapted for MARL. 

    \begin{figure}[t]	
    \vspace{.5em}
    \begin{center}
    	\includegraphics[width=.85\linewidth]{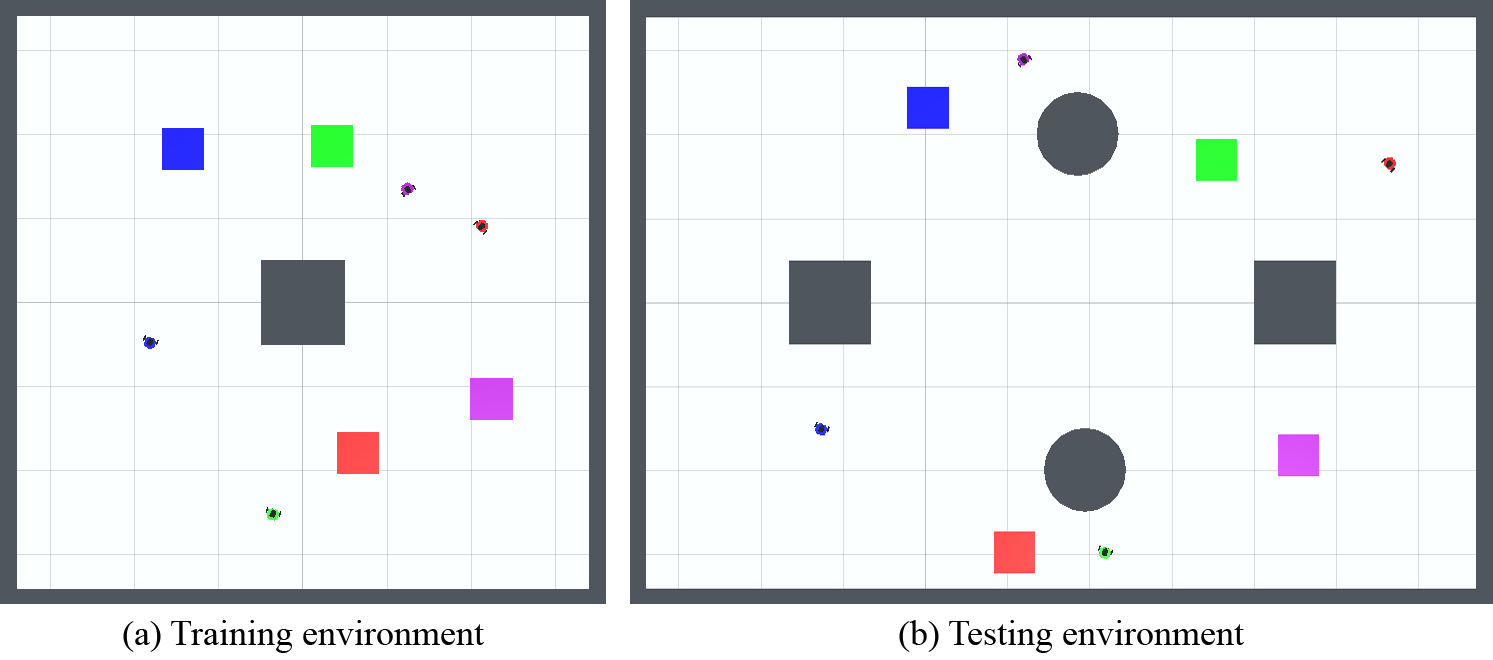}
        \caption{Left: Training environment for our MARL navigation. Right: Testing environment for the trained policies. Explanatory view with $N = 4$ robots in different colors.}
        \label{fig:environments}
    \end{center}
    \vspace{-1em}
    \end{figure}
    
In more detail, we consider a Turtlebot3 indoor navigation environment with $N \in [2, 4, 8]$ robots and fixed obstacles, where each robot has to navigate to its target avoiding collisions. 
Given the robot specifications, we consider a max angular velocity of $90$ $deg/s$ and a max linear velocity $= 0.2$ $m/s$. The decision-making frequency of the robot in both the training and testing environments is set to 20Hz, to reflect the update rate of the equipped LDS-01 lidar sensor.
This training environment is depicted in Fig. \ref{fig:environments}a and is built with the Unity ml-agents toolkit \cite{unity}. As demonstrated in \cite{drl_navigation2}, Unity enables rapid development of simulation environments, which also allows the export of trained models to the more conventional Robot Operating System (ROS). To detect collisions among the robots and the obstacles, we used Unity's mesh collider system, applied to the 3D models of the environment. This allows triggering a collision event when an intersection between the bounds of two or more colliders is detected. 
The target goals of each randomly spawn in the scenario and are guaranteed to be obstacle-free and with a minimum distance between each other (set to $0.5m$).
Each agent receive a separate reward $r_t$ that is structured as follows:

\begin{equation}
    r(s_t) =
    \begin{cases}
        \alpha (d_{t-1} - d_t) - 0.005, \text{ if } d_t > 0.1 \\
        1, \text{ if } d_t <= 0.1 \\
        -1, \text{ if a collision is detected} 
    \end{cases}
\end{equation}

\noindent hence we have two sparse value in case of reaching the target within a threshold distance $d_t = 10cm$ from the robot's goal (i.e., $r_t = 1$), or crashing within the obstacles or other robots (i.e., $r_t = -1$) which terminates an episode resetting the robot to its starting position and generating a new set of goals. A dense component is used during the travel: $\alpha(d_{t-1} - d_t) - 0.005$, where $d_{t-1}$, $d_t$ is euclidean distance between a robot and its goal at two consecutive time steps and $\alpha$ is a multiplicative factor that is used for normalization. A penalty of $0.005$ is applied at each time-step, to encourage the robots to perform the shortest path possible.

\subsubsection{Network Architecture}
\begin{figure}[t]	
    \vspace{.5em}
    \begin{center}
    	\includegraphics[width=.9\linewidth]{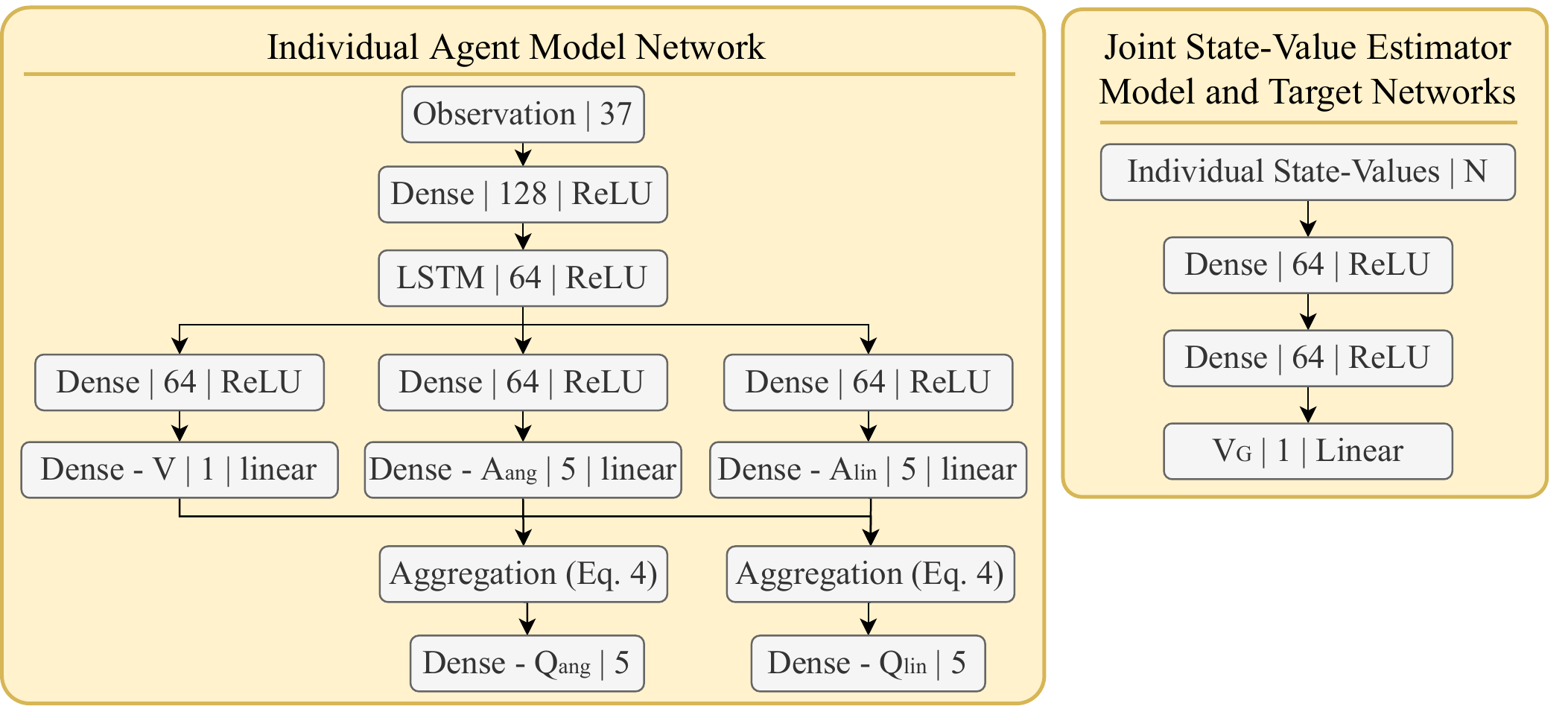}
        \caption{Left: robot network architecture. Right: join state-value estimator network architecture}
        \label{fig:networks}
    \end{center}
    \vspace{-1em}
    \end{figure}

The input layer of each network contains a 35-dimensional vector with sparse laser scan values sampled in $[-120, 120]$ degrees in a fixed angle distribution, and the individual target position is expressed as the distance from the robot and relative heading. We considered a similar setting of previous DRL literature for robotic navigation \cite{drl_navigation2, drl_navigation3}.
The output layer considers the Branching Architecture \cite{branching} to output two streams of $Q$-values: one for angular velocities $v_{ang} \in [-90, -45, 0, 45, 90]~deg/s$ and one for linear velocities $v_{lin} \in [0, 0.05, 0.1, 0.15, 0.2]~m/s$. Hence, following the GDQ architecture described in Section \ref{sec:methods}, each network maintains three separate streams: one for the advantage of $v_{ang}$, one for the advantage of $v_{lin}$, and one to estimate the state-value. The two advantage streams are then combined with the state-value using Eq. \ref{eq:duelingq}. Given the importance of the statistical significance of the data, highlighted by several recent work \cite{eval1, eval2}, we performed an initial evaluation in the training environment with multiple trials on different network sizes and seeds. The outcome led us to use the network architectures depicted in Fig. \ref{fig:networks}, with one \textit{ReLU} hidden layer with $128$ neurons, a \textit{ReLU} LSTM layer of size $64$, plus an additional \textit{ReLU} hidden layer of $64$ neurons for each stream. The joint state-value estimator, is a simple feed-forward network with two \textit{ReLU} hidden layers of $64$ neurons each and a \textit{linear} output.

\subsection{Empirical Evaluation}
Given the results in our preliminary evaluation in Section \ref{subsec:example}, we evaluate the performance of the two best-performing algorithms for our multi-robot mapless navigation: QMIX, and the proposed GDQ. The goal of this evaluation is to investigate whether GDQ can successfully address a robotic task of real interest while favoring cooperation as detailed in Section \ref{sec:methods}.
Data are collected on an i7-9700k, using a network architecture for QMIX similar to the one described in the previous section in terms of layer numbers and sizes (while we refer the interested reader to the original work to further details about QMIX \textit{hypernetworks} \cite{qmix}). Every experiment with $N$ robots $\in [2, 4, 8]$ is performed over ten different independent runs with different seeds, for statistical significance of the collected data. For each experiment, we report the following curves that show mean and standard deviation over the runs, smoothed over $100$ epochs. In detail, for each setting, we show the average success rate: how many successful obstacles-free trajectories are performed over a sequence of $100$ epochs. In the evaluation in an unseen scenario (Tab. \ref{tab:evaluation} considers average reward: the cumulative reward for each epoch, and average path length: represented as the number of steps in the environment for each epoch. All these metrics are averaged over the $N$ robots.

\subsubsection{Training Results}
\begin{figure*}[t]	
    \vspace{.5em}
	\centering
	\begin{subfigure}[t]{0.2\linewidth}
		\centering
		\includegraphics[width=1\linewidth]{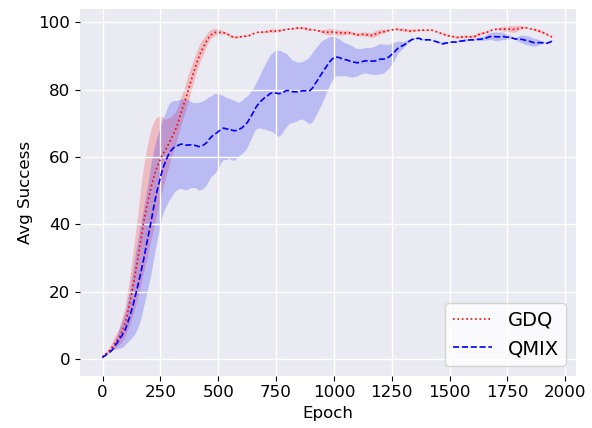}
		\caption{$N = 2$}
		\label{fig:2agents}
	\end{subfigure}
	\quad
	\begin{subfigure}[t]{0.2\linewidth}
		\centering
		\includegraphics[width=1\linewidth]{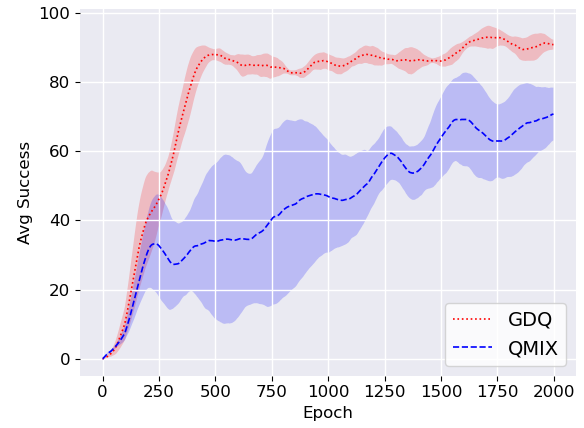}
		\caption{$N = 4$}
		\label{fig:4agents}
	\end{subfigure}
	\begin{subfigure}[t]{0.2\linewidth}
		\centering
		\includegraphics[width=1\linewidth]{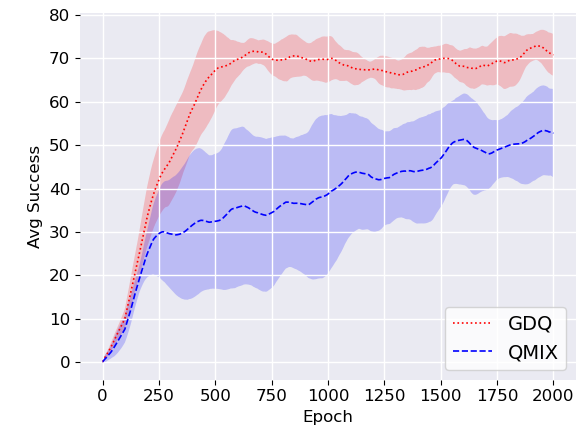}
		\caption{$N = 8$}
		\label{fig:8agents}
	\end{subfigure}
	\caption{Average Success Rate of the GDQ and QMIX robots in our training phase.}
	\label{fig:graph_results}
\end{figure*}

Fig. \ref{fig:graph_results} shows the results of GDQ and QMIX in our robotic navigation tasks, where the robots have to learn how to reach different target positions, avoiding collisions with the fixed obstacles and with each other.

For each task with $N \in [2, 4, 8]$ Turtlebot3 robots we show the average success rate. While the results are comparable when considering only 2 robots (i.e., Fig. \ref{fig:2agents}), with a growing number of robots GDQ clearly offers better performances. In more detail, with $N = 4$ (i.e., Fig. \ref{fig:4agents}) the percentage of successes stabilizes at over $\approx$90\% in $\approx$1700 epochs, i.e., $150$ minutes of training. QMIX, in contrast, reaches $\approx$80\% successes in the same epochs that correspond to $190$ minutes of training. It naturally follows that GDQ offers better performance even in terms of reward and, as shown in the following evaluation, it performs on average a lower number of steps, which means shorter paths. The evaluation with $N = 8$ confirms the superior performance of GDQ.

\subsubsection{Testing Results}
Crucially, after the training phase, the converged models can navigate in the environment in a decentralized fashion, exploiting the minimal information in the input layer of the network. Moreover, the individuals generalize crucial aspects of robot navigation such as (i) robot starting position, (ii) target position, (iii) velocity. The laser scan-based navigation also allows our robots to navigate in previously unknown environments with different obstacles, which is a key feature for motion planning. To confirm the successful generalization of our models, we run an additional evaluation with $N \in [4, 8]$ in a previously unseen scenario, depicted on the right Fig. \ref{fig:environments}. Results in Tab. \ref{tab:evaluation} confirm the trend highlighted during the training, where GDQ offers superior performance over the baseline in terms of average reward and path length when the same sequence of goals is considered for both algorithms. In this evaluation, we also collect the percentage of collisions detected, which further confirms a better generalization of the task with multiple robots in the case of GDQ.

\begin{table}[t]
    \caption{Average Reward, Steps and Collision percentage in the evaluation over ten runs on different targets in the unseen scenario depicted in the right Fig. \ref{fig:environments}}
    \label{tab:evaluation}
    \centering
    \begin{tabular}{lllll}
    \toprule
    & \textbf{N} & \multicolumn{1}{c}{\textbf{Reward}} & \multicolumn{1}{c}{\textbf{Steps}} & \multicolumn{1}{c}{\textbf{Collisions (\%)}} \\ 
    \midrule
    \textbf{QMIX} & 4 & 26.3 $\pm$ 3.1 & 295 $\pm$ 23 & 1.4 $\pm$ 2.6   \\
    \textbf{GDQ}  & 4 & 30.1 $\pm$ 2.7 & 242 $\pm$ 18 & 0.8 $\pm$ 1.8 \\ 
    \midrule
    \textbf{QMIX} & 8 & 18.2 $\pm$ 3.9 & 361 $\pm$ 29 & 34.2 $\pm$ 4.1   \\
    \textbf{GDQ}  & 8 & 24.3 $\pm$ 2.8 & 285 $\pm$ 22 & 17.3 $\pm$ 2.3  \\ 
    \bottomrule
    \end{tabular}
\end{table}

\section{Discussion}
\label{sec:discussion}

We presented GDQ, a novel value-based CTDE approach for multi-agent scenarios that exploit state-values information to favor cooperation. We performed an initial evaluation of GDQ and several baselines (i.e., IQL, VDN, QMIX) in the \textit{Cooperative Navigation} benchmark to highlights the superior performance of our approach. Then, we presented a multi-robot robotic navigation scenario built in Unity to evaluate the navigation performance of GDQ with 2, 4, and 8 robots. 

Our empirical evaluation shows that GDQ significantly outperforms prior CTDE approaches in multi-robot robotic navigation, especially with a growing number of robots. Crucially, such superior performance is obtained only considering the individual's state-values and no additional centralized information (such as in QMIX), which allows scaling the number of robots without major overhead for the training process (each robot only adds an input node in our centralized joint state-value estimator). 

This work paves the way for several interesting research directions which include the exploration of competitive multi-agent tasks using separate joint state-value estimators.


\bibliographystyle{IEEEtran}
\bibliography{root.bib}

\begin{thebibliography}{10}
\providecommand{\url}[1]{#1}
\csname url@samestyle\endcsname
\providecommand{\newblock}{\relax}
\providecommand{\bibinfo}[2]{#2}
\providecommand{\BIBentrySTDinterwordspacing}{\spaceskip=0pt\relax}
\providecommand{\BIBentryALTinterwordstretchfactor}{4}
\providecommand{\BIBentryALTinterwordspacing}{\spaceskip=\fontdimen2\font plus
\BIBentryALTinterwordstretchfactor\fontdimen3\font minus
  \fontdimen4\font\relax}
\providecommand{\BIBforeignlanguage}[2]{{%
\expandafter\ifx\csname l@#1\endcsname\relax
\typeout{** WARNING: IEEEtran.bst: No hyphenation pattern has been}%
\typeout{** loaded for the language `#1'. Using the pattern for}%
\typeout{** the default language instead.}%
\else
\language=\csname l@#1\endcsname
\fi
#2}}
\providecommand{\BIBdecl}{\relax}
\BIBdecl

\bibitem{marl_coopeartive}
K.~Tuyls and G.~Weiss, ``Multiagent learning: Basics, challenges, and
  prospects,'' in \emph{AI Magazine}, 2012.

\bibitem{mr_search}
J.~L. Baxter, E.~K. Burke, J.~M. Garibaldi, and M.~Norman, ``Multi-robot search
  and rescue: A potential field based approach,'' in \emph{Autonomous Robots
  and Agents}, 2007.

\bibitem{mr_collisionavoidance}
P.~Long, T.~Fan, X.~Liao, W.~Liu, H.~Zhang, and J.~Pan, ``Towards optimally
  decentralized multi-robot collision avoidance via deep reinforcement
  learning,'' in \emph{ICRA}, 2018.

\bibitem{vdn}
P.~Sunehag, G.~Lever, A.~Gruslys, W.~M. Czarnecki, V.~F. Zambaldi,
  M.~Jaderberg, M.~Lanctot, N.~Sonnerat, J.~Z. Leibo, K.~Tuyls, and T.~Graepel,
  ``Value-decomposition networks for cooperative multi-agent learning,'' in
  \emph{AAMAS}, 2018.

\bibitem{maddpg}
R.~Lowe, Y.~Wu, A.~Tamar, J.~Harb, P.~Abbeel, and I.~Mordatch, ``Multi-agent
  actor-critic for mixed cooperative-competitive environments,'' in
  \emph{NIPS}, 2017.

\bibitem{qmix}
T.~Rashid, M.~Samvelyan, C.~S. de~Witt, G.~Farquhar, J.~N. Foerster, and
  S.~Whiteson, ``{QMIX:} monotonic value function factorisation for deep
  multi-agent reinforcement learning,'' in \emph{ICML}, 2018.

\bibitem{qtran}
K.~Son, D.~Kim, W.~J. Kang, D.~Hostallero, and Y.~Yi, ``{QTRAN:} learning to
  factorize with transformation for cooperative multi-agent reinforcement
  learning,'' in \emph{ICML}, 2019.

\bibitem{particleenv}
I.~Mordatch and P.~Abbeel, ``Emergence of grounded compositional language in
  multi-agent populations,'' in \emph{arXiv}, 2017.

\bibitem{duelingdqn}
Z.~Wang, N.~de~Freitas, and M.~Lanctot, ``Dueling network architectures for
  deep reinforcement learning,'' in \emph{ICML}, 2016.

\bibitem{doubledqn}
H.~van Hasselt, A.~Guez, and D.~Silver, ``Deep reinforcement learning with
  double q-learning,'' in \emph{AAAI}, 2016.

\bibitem{decpomdp}
F.~A. Oliehoek and C.~Amato, ``A concise introduction to decentralized
  pomdps,'' in \emph{SpringerBriefs in Intellingece Systems}, 2016.

\bibitem{eumaspomcp}
A.~Castellini, E.~Marchesini, G.~Mazzi, and A.~Farinelli, ``Explaining the
  influence of prior knowledge on pomcp policies,'' in \emph{Multi-Agent
  Systems and Agreement Technologies}, 2020.

\bibitem{per}
T.~Schaul, J.~Quan, I.~Antonoglou, and D.~Silver, ``Prioritized experience
  replay,'' in \emph{ICLR}, 2016.

\bibitem{rainbow}
M.~Hessel, J.~Modayil, H.~Van~Hasselt, T.~Schaul, G.~Ostrovski, W.~Dabney,
  D.~Horgan, B.~Piot, M.~Azar, and D.~Silver, ``Rainbow: Combining improvements
  in drl,'' in \emph{AAAI}, 2018.

\bibitem{drl_navigation1}
J.~{Zhang}, J.~T. {Springenberg}, J.~{Boedecker}, and W.~{Burgard}, ``Deep
  reinforcement learning with successor features for navigation across similar
  environments,'' in \emph{IROS}, 2017.

\bibitem{drl_navigation3}
L.~Tai, G.~Paolo, and M.~Liu, ``Virtual-to-real drl: Continuous control of
  mobile robots for mapless navigation,'' in \emph{IROS}, 2017.

\bibitem{prove}
D.~Corsi, E.~Marchesini, and A.~Farinelli, ``Formal verification of neural
  networks for safety-critical tasks in deep reinforcement learning,'' in
  \emph{UAI}, 2021.

\bibitem{superl}
E.~Marchesini, D.~Corsi, and A.~Farinelli, ``Genetic soft updates for policy
  evolution in deep reinforcement learning,'' in \emph{ICLR}, 2021.

\bibitem{drl_navigation2}
E.~{Marchesini} and A.~{Farinelli}, ``Discrete deep reinforcement learning for
  mapless navigation,'' in \emph{ICRA}, 2020.

\bibitem{mixed_gdrl}
E.~Marchesini and A.~Farinelli, ``Genetic deep reinforcement learning for
  mapless navigation,'' in \emph{AAMAS}, 2020.

\bibitem{branching}
A.~Tavakoli, F.~Pardo, and P.~Kormushev, ``Action branching architectures for
  deep reinforcement learning,'' in \emph{AAAI}, 2018.

\bibitem{mujoco}
E.~Todorov, T.~Erez, and Y.~Tassa, ``Mujoco: A physics engine for model-based
  control,'' in \emph{IROS}, 2012.

\bibitem{valuebased1}
T.~Wiele, D.~Warde-Farley, A.~Mnih, and V.~Mnih, ``Q-learning in enormous
  action spaces via amortized approximate maximization,'' in \emph{NeurIPS
  Workshop}, 2018.

\bibitem{valuebased2}
G.~Dulac{-}Arnold, R.~Evans, P.~Sunehag, and B.~Coppin, ``Reinforcement
  learning in large discrete action spaces,'' in \emph{CoRR}, 2015.

\bibitem{coma}
J.~N. Foerster, G.~Farquhar, T.~Afouras, N.~Nardelli, and S.~Whiteson,
  ``Counterfactual multi-agent policy gradients,'' in \emph{AAAI}, 2018.

\bibitem{poor_factorization}
A.~Mahajan, T.~Rashid, M.~Samvelyan, and S.~Whiteson, ``{MAVEN:} multi-agent
  variational exploration,'' in \emph{NeuIPS}, 2019.

\bibitem{hypernetworks}
D.~Ha, A.~M. Dai, and Q.~V. Le, ``Hypernetworks,'' in \emph{ICLR}, 2017.

\bibitem{mr_selfdriving}
P.~Wang, C.~Chan, and A.~de~La~Fortelle, ``A reinforcement learning based
  approach for automated lane change maneuvers,'' in \emph{IEEE Intelligent
  Vehicles Symposium}, 2018.

\bibitem{unity}
A.~Juliani, V.~Berges, E.~Vckay, Y.~Gao, H.~Henry, M.~Mattar, and D.~Lange,
  ``Unity: A platform for intelligent agents,'' in \emph{CoRR}, 2018.

\bibitem{eval1}
C.~Colas, O.~Sigaud, and P.-Y. Oudeyer, ``A hitchhiker's guide to statistical
  comparisons of rl algorithms,'' in \emph{arXiv}, 2019.

\bibitem{eval2}
P.~Henderson, R.~Islam, P.~Bachman, J.~Pineau, D.~Precup, and D.~Meger, ``Deep
  reinforcement learning that matters,'' in \emph{AAAI}, 2018.

\end{thebibliography}


\end{document}